\documentclass[sigconf]{acmart}

\AtBeginDocument{%
  \providecommand\BibTeX{{%
    \normalfont B\kern-0.5em{\scshape i\kern-0.25em b}\kern-0.8em\TeX}}}

\setcopyright{acmcopyright}
\copyrightyear{2022} 
\acmYear{2022} 
\setcopyright{acmcopyright}
\acmConference[RecSys '22]{Sixteenth ACM Conference on Recommender Systems}{September 18--23, 2022}{Seattle, WA, USA}
\acmBooktitle{Sixteenth ACM Conference on Recommender Systems (RecSys '22), September 18--23, 2022, Seattle, WA, USA}
\acmPrice{15.00}
\acmDOI{10.1145/3523227.3546768}
\acmISBN{978-1-4503-9278-5/22/09}

\acmSubmissionID{1088}

\usepackage{arydshln}
\usepackage{graphicx}
\usepackage{enumitem}
\usepackage{acmart-taps}
\usepackage[labelfont=bf]{caption}

\begin{document}

\title[Exploiting Negative Preference in Content-based Music Recommendation]{Exploiting Negative Preference in Content-based Music Recommendation with Contrastive Learning}

\author{Minju Park}
\email{minju0821@snu.ac.kr}
\affiliation{%
  \institution{Department of Intelligence and Information \\ Seoul National University}
  \city{Seoul}
  \country{Republic of Korea}}

\author{Kyogu Lee}
\email{kglee@snu.ac.kr}
\affiliation{%
  \institution{Department of Intelligence and Information}
  \institution{Graduate School of AI}
  \institution{AI Institute \\ Seoul National University}
  \city{Seoul}
  \country{Republic of Korea}}

\begin{abstract}
Advanced music recommendation systems are being introduced along with the development of machine learning.
However, it is essential to design a music recommendation system that can increase user satisfaction by understanding users' music tastes, not by the complexity of models.
Although several studies related to music recommendation systems exploiting negative preferences have shown performance improvements, there was a lack of explanation on how they led to better recommendations.
In this work, we analyze the role of negative preference in users' music tastes by comparing music recommendation models with contrastive learning exploiting preference (CLEP) but with three different training strategies - exploiting preferences of both positive and negative (CLEP-PN), positive only (CLEP-P), and negative only (CLEP-N).
We evaluate the effectiveness of the negative preference by validating each system with a small amount of personalized data obtained via survey and further illuminate the possibility of exploiting negative preference in music recommendations.
Our experimental results show that CLEP-N outperforms the other two in accuracy and false positive rate.
Furthermore, the proposed training strategies produced a consistent tendency regardless of different types of front-end musical feature extractors, proving the stability of the proposed method.
\end{abstract}



\maketitle

\section{Introduction}

The popularization of digital music streaming services has made a lot of music accessible to people.
As a result, personalized music recommendation technology has become an essential factor for both users and online music streaming services.
The need and interest in music recommendations have increased, and many related studies have been conducted actively in recent years.
Various hybrid recommendation methods have been proposed, led by the collaborative filtering methods in which recommendations are made based on the user's listening history and the content-based filtering methods which are based on the song's content.

However, recent studies on music recommendation systems were mainly conducted to improve performance by adding new features or using new machine learning techniques such as latent factor models or deep representation learning ~\cite{schedl2019deep}.
There is no doubt that improving the model's performance is essential, but fundamental analysis of why better recommendations have become possible is sometimes overlooked due to the focus on the engineering perspective.
~\cite{zhang2020explainable} pointed out that recent advances in more complex recommendation models further brought the difficulty of transparency.
Recommendation is all about personalization, and its objective is to model one's music taste.
It is supported by ~\cite{koenigstein2017rethinking}, indicating that the actual goal of most real-world recommendation systems is "to influence the user to consume more items than she would have without the recommendations, not to predict the next item the user will consume."
In order to systematically explain the mechanisms of improvement in recommendation systems, one's music taste ought to be understood.

The naïve idea that motivated our work is, “Isn't it easier to explain the music I hate than the music I like?”
Several studies have shown better performance when implementing negative feedback in recommendation systems.
~\cite{chao2005adaptive, pampalk2005dynamic} applied negative feedback in music recommendations, but they focused on proposing new architecture designs considering the negative feedback as an additional feature.
Our work goes beyond simply proposing a recommendation system to which negative feedback is applied and aims to illustrate the role of negative feedback in modeling music taste.
In a way, one's music taste can be seen as a set of pairs consisting of songs and corresponding preferences.
Thus the song's content must be considered in order to approach the concept of music taste.

Nevertheless, there has been no attempt to explain music taste by applying positive and negative feedback to content-based music recommendation systems.
In our work, we will apply negative feedback based on the content-based filtering method and explain some parts of music taste focusing on negative feedback.
To prevent the ambiguity of negative feedback and naturally relate it to the concept of music taste, we will use the term "negative preference" by borrowing the expression of ~\cite{chao2005adaptive}.

We introduce three content-based music recommendation systems with differently conditioned contrastive learning exploiting preference (CLEP), designed based on Siamese Neural Network (SNN) ~\cite{koch2015siamese}.
The three models differ in the process of computing the final embedding vectors of the songs according to the targeted preferences - model exploiting both positive and negative preferences (CLEP-PN), model exploiting positive preference only (CLEP-P), and model exploiting negative preference only (CLEP-N).
CLEP-PN embeds songs in a way that both positive and negative preferences are characterized.
CLEP-P embeds songs in a way that positive preference is solely characterized, and CLEP-N embeds songs in a way that negative preference is solely characterized.
Three different representations for each song will then be obtained from the frozen networks and will be used to train a simple classifier to match the preferences of each song.
Afterward, the models are trained to fit a single user to fully analyze the effect of personal preferences.
We generated a user preference dataset via survey, consisting of pairs of songs and corresponding preferences for every user, obtained from twenty-four participants.
The models are then trained with each dataset to predict the participant's preference for a new song.
For the training, the features of each song are represented using existing works of musical feature extraction.
To guarantee the stability of our work, we used three different musical feature extraction methods - Contrastive Learning of Musical Representations (CLMR) ~\cite{spijkervet2021contrastive}, Music Effects Encoder (MEE) ~\cite{koo2022end}, and Jukebox ~\cite{dhariwal2020jukebox}.
Our models will finally be evaluated on the test set with accuracy, precision, recall, area under the receiver operating characteristic curve (AUROC), and false positive rate.
By comparing these metrics of the three models, we will be able to understand the effects of preferences, especially negative preferences, and further explain the relevant parts of music taste.

Throughout our work, we will be investigating the following research questions.
\begin{itemize}
\item \textbf{RQ 1. What characteristics do negative preferences have in terms of explaining music taste?}

We will identify that compared to positive preference, negative preference in music taste has more distinct characteristics and that it is easier to explain music taste through negative preference. \\

\item \textbf{RQ 2. How does applying negative preference help improve music recommendations?}

We will discuss the advantages of exploiting negative preference in music recommendations through the identified roles of negative preference in music taste.

\end{itemize}

\section{Related Works}

In this section, we provide an overview of content-based music recommendations and previous attempts to exploit negative preferences in recommendation systems. 

\subsection{Content-based Music Recommendation}

Content-based music recommendation systems have a strong advantage in that the audio content itself is utilized.
Due to its reliance on the content, it compensates for the limitations of collaborative filtering methods, such as the cold-start problem, which is a problem caused by a deficiency in the information about new items or new users. 
Traditional content-based music recommendation systems are mainly based on metadata such as artists, albums, or genres.
However, developments in music information retrieval have facilitated the handling of music content in various ways.
It became possible for both high-level audio features (e.g., melody, harmony, rhythm) and low-level audio features (e.g., Mel-Frequency Cepstral Coefficients (MFCC), mel-spectrogram) to be used for music representation.
Furthermore, musical feature extractors utilizing advanced deep learning techniques have also been proposed.
~\cite{spijkervet2021contrastive} and ~\cite{koo2022end} use the idea of contrastive learning, and ~\cite{dhariwal2020jukebox} use the idea of multi-scale VQ-VAE to extract low-level features of music.

Accordingly, content-based music recommendations adopting these features are being suggested, demonstrating their practical applicability ~\cite{deldjoo2021content}.
In addition to using various methods in extracting musical features, deep learning techniques based on simple front-end features are used to predict music's latent factors which can be utilized for content-based music recommendations.
~\cite{van2013deep} proposed a latent factor model which maps mel-spectrogram to the item latent factor vectors obtained from the collaborative filtering method using deep convolutional neural networks.
Based on different representations, content-based music recommendation systems compute the distances between songs and recommend songs similar to the ones the user likes.
The computation methods of distance also vary by model ~\cite{liu2013comparison}, but the point to note is that content-based music recommendation systems usually rely on similarity.
High reliance on similarity causes the recommendations to lack novelty, and content similarity was once criticized for not being able to completely capture the preferences of a user ~\cite{kaminskas2012contextual}.
We expect to overcome these problems by exploiting user preference data along with the contents, referring to the work of ~\cite{van2013deep} which successfully bridges the semantic gap in the content-based filtering method by using the ground truth, which includes user feedback information.

\vspace*{-5pt}
\subsection{Recommendation Systems Exploiting Negative Preference}

Recommendation systems rely on user feedback, which can be divided into explicit feedback (e.g., ratings) and implicit feedback (e.g., browsing history, purchase history) according to how it is provided.
Implicit feedback outnumbers explicit feedback due to its continuous update, but unfortunately, implicit feedback has a constraint that it is mainly focused on positive feedback  ~\cite{kelly2003implicit, gauch2007user}.
Thus modern recommendation systems are predominantly based on positive feedback, followed by the concern of its deficiencies in discriminatory power ~\cite{hu2008collaborative, lu2018between}.
In this regard, several studies are attempting to exploit negative feedback, or "negative preference" in our term, in recommendation systems ~\cite{chao2005adaptive, pampalk2005dynamic}.
~\cite{chao2005adaptive} applied negative preference in group recommendation, showing that negative preference helped groups find consensus solutions satisfactory to all individuals.
They introduced a recommendation system of avoiding the item user does not want rather than recommending the item user wants and raised the possibility of applying negative preference in recommendations.
Recommendation systems using positive and negative preferences were proposed in various domains, applying different learning models.
For instance, ~\cite{clements2009exploiting, chen2013modified} exploited both positive and negative preferences by modifying graph-based recommendation systems.
Proposed models have commonly shown that the user's negative preferences have increased the quality of recommendations ~\cite{lee2009reinforcing, chen2021movie}.
For music recommendations, negative preferences are applied through skipping behaviors.
~\cite{pampalk2005dynamic} introduced a heuristic of automatic playlist generation by eliminating songs similar to the skipped songs.
Studies covering sequential skip prediction tasks ~\cite{ferraro2019skip, montecchio2020skipping, chang2019sequential} also imply the possibilities of exploitation of negative preference in music recommendations.

Research in this field constantly mentions negative preferences, but most conclude by proposing novel architecture designs with increased performance.
In contrast to these related works, we expect to focus on illuminating the specific roles of negative preferences compared to positive preferences.

\section{Methods}

\begin{figure}
    \includegraphics[width=\columnwidth]{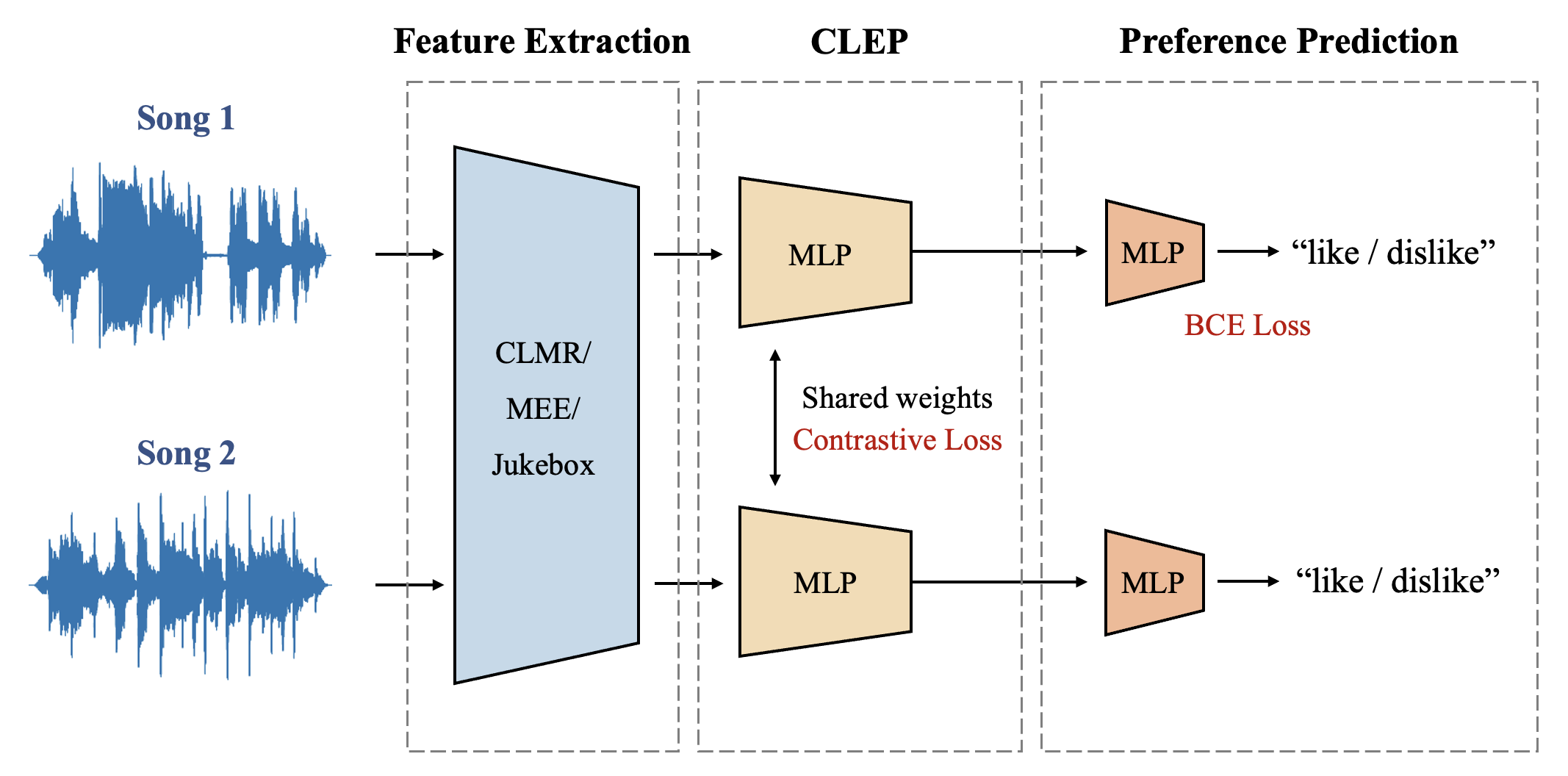}
    \caption{\textbf{Overview of the proposed method}}
    \label{fig:model}
\end{figure}

The main goal of our study is to understand the effects of negative preference through a comparison of recommendation models in which preferences are differently conditioned.
For methodical investigation, we designed the framework to which our model will be applied consisting of three parts: feature extraction, embedding with CLEP, and preference prediction.
The framework overview is visualized in Figure \ref{fig:model}.

\subsection{Feature Extraction}

Extracting the features of songs is essential in content-based music recommendations.
As described in Section 2.1, previous studies have introduced content-based music recommendation systems using various features.
Considering that our work attempts to relate the contents with user feedback, we use low-level features following ~\cite{van2013deep}.
As representation learning has been actively studied in recent years, ~\cite{spijkervet2021contrastive, koo2022end, dhariwal2020jukebox} proposed models trained in a self-supervised manner that produce novel music representations.
There are also front-end models used in automatic music tagging ~\cite{pons2019musicnn, choi2017convolutional}, but they are models trained to classify music into a limited, discrete range of descriptions.
These tags can help express the music users accept, but a much more intricate approach to representation is required to relate to their music preferences.
Therefore, rather than tag-based models, models based on self-supervised contrastive learning can be considered appropriate for this study since it is trained to focus on the identity of the music content itself.

By taking advantage of the pre-trained self-supervised models, the training procedure can be eased as we only need to train back-end models using a small amount of preference data of a single person.
We use different feature extractors to evaluate the stability of our proposed method despite its varying performance according to each music representation.
Our work uses the framework of CLMR ~\cite{spijkervet2021contrastive}, MEE ~\cite{koo2022end}, and Jukebox ~\cite{dhariwal2020jukebox} as front-end musical feature extractors. CLMR is a method of extracting musical features based on the idea of SimCLR ~\cite{chen2020simple}, which performs contrastive learning by designating different sections of the same song as positive samples and sections of different songs as negative samples.
MEE modifies the idea of CLMR to capture the song's overall timbre and mood, including its mastering style.
Jukebox introduces Music VQ-VAE using the architecture of hierarchical VQ-VAE ~\cite{van2017neural, razavi2019generating}, and its encoder successfully represents the music with latent vectors.
Thanks to the provision of pre-trained models, we take each model to extract the features of songs for our work.
Details of the musical feature extraction models are illustrated in Table \ref{tab:data}.
The amount of data was too small to show statistically significant results by training the front-end model from scratch, and it was shown as we expected in preliminary experiments - training a simple convolutional neural network (CNN) with mel-spectrogram input and training a network adopting the idea of CLMR.
Therefore, we will be focusing on the pre-trained musical feature extractors for the rest of our work.

\begin{table}[]
\caption{\textbf{Details of the front-end musical feature extraction models}}
\resizebox{\columnwidth}{!}{
\renewcommand{\arraystretch}{1.2}
\begin{tabular}{cccc}
\hline
\textbf{Front-end Models} & \textbf{Dataset (\# Tracks)} & \begin{tabular}[c]{@{}c@{}}\textbf{Sampling Rate,}\\ \textbf{Channel}\end{tabular} & \textbf{Dimension}\\ \hline
CLMR & MagnaTagATune ~\cite{law2009evaluation} (187k) & 16 kHz, mono & 512 \\
MEE & MTG-Jamendo ~\cite{bogdanov2019mtg} (55k) & 44.1kHz, stereo & 2048 \\
Jukebox & web crawled (1.2m) & 44.1kHz, mono & 4800 \\ \hline
\end{tabular}
}
\label{tab:data}
\end{table}
\vspace{-5pt}

\begin{figure*}
    \includegraphics[width=1.9\columnwidth]{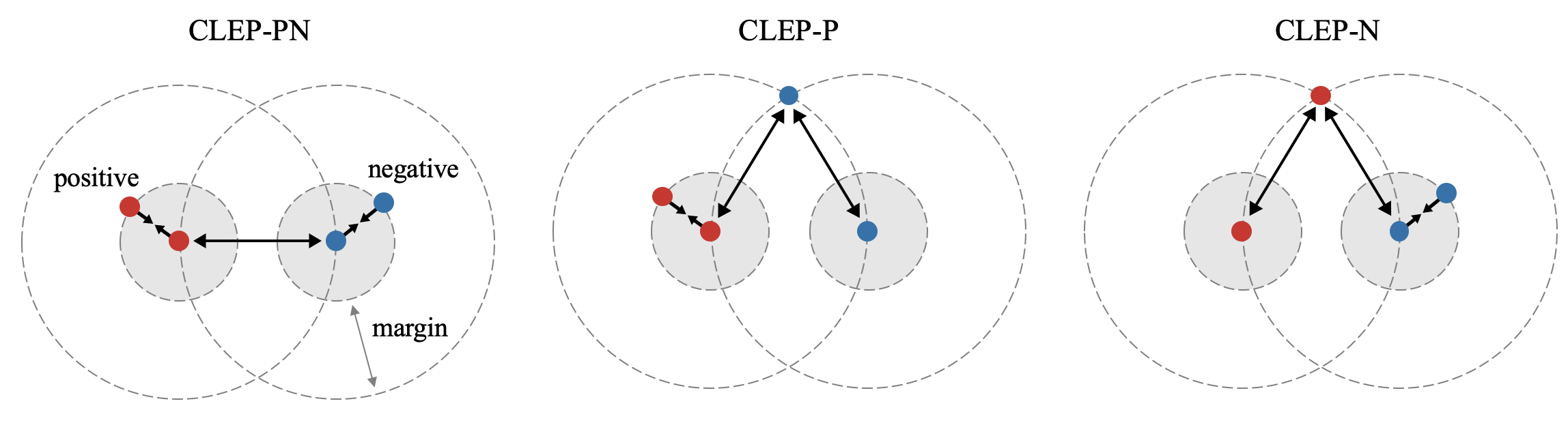}
    \caption{\textbf{Demonstration of each embedding space of CLEP-PN, CLEP-P, and CLEP-N}}
    \label{fig:clep}
\end{figure*}

\vspace*{10pt}
\subsection{Contrastive Learning Exploiting Preference (CLEP)}

We devise three different content-based music recommendation models as follows:
\begin{itemize}
\item \textbf{CLEP-PN} \\
Model with contrastive learning exploiting both positive and negative preferences
\item \textbf{CLEP-P} \\
Model with contrastive learning exploiting positive preference only
\item \textbf{CLEP-N} \\
Model with contrastive learning exploiting negative preference only
\end{itemize}

The three models are differentiated in the embedding part.
The representations obtained in the previous part are embedded considering the preferences using the architecture of SNN.
SNN learns representations by adjusting the distance between the embeddings according to the labels of item pairs.
In more detail, SNN is trained with contrastive loss as follows: 
\begin{equation}
L^{Contrastive} = yD^2 + (1-y)max(margin-D,0)^2
\end{equation}
where $y$ is the label of an item pair and $D$ is the distance between the items.
When a pair of items labeled as $y=1$ is given, it leads to $L=D^2$, reducing the distance as training.
That is, a pair of items that is labeled as $y=1$ will be embedded close together in the embedding space.
On the other hand, when a pair of items labeled as $y=0$ is given, it leads to $L=max(margin-D,0)^2$.
So as training continues, the distance gets close to the margin value.
The margin value was set as $margin=7$ through empirical observations so that the embeddings from both classes were well separated. 

As it can be seen from the loss function, the embedding varies depending on how the item pairs are labeled.
In general classification tasks, items that belong to the same class are embedded closer and those belonging to different classes are embedded farther. 
We changed the way of labeling according to the purpose of each model like the following, and the embedding strategies of each model are depicted in Figure \ref{fig:clep}.

\begin{itemize}
\item \textbf{CLEP-PN}

The label is set so that the songs with the same preference are embedded close to each other, and the songs with different preferences are embedded far apart.
In other words, we set $y=1$ for 'like-like' and 'dislike-dislike' pairs, and $y=0$ for 'like-dislike' pairs.

\item \textbf{CLEP-P}

The label is set so that the songs with positive preferences are embedded close together, and other kinds of pairs are embedded far apart.
We set $y=1$ for 'like-like' pairs, and $y=0$ for 'like-dislike' and 'dislike-dislike' pairs.

\item \textbf{CLEP-N}

The label is set so that the songs with negative preferences are embedded close together, and other kinds of pairs are embedded far apart.
We set $y=1$ for 'dislike-dislike' pairs, and $y=0$ for 'like-dislike' and 'like-like' pairs.
\end{itemize}

\subsection{Preference Prediction}

Pre-trained musical feature extractors are often evaluated in classification tasks by appending a simple model of Multi-Layer Perceptron (MLP) ~\cite{spijkervet2021contrastive}.
We apply the same technique to predict the user's preference for each song.
MLP layers are added and trained to match the ground truth of whether the user likes or dislikes the song, with Binary Cross Entropy loss (BCE loss).
Then the sigmoid function eventually computes the probability of preference.

\section{Experiments}

\subsection{Experimental Setups}

For musical feature extractors, we used the public-available pre-trained models of CLMR \footnote{https://github.com/Spijkervet/CLMR}, MEE  \footnote{https://github.com/jhtonyKoo/e2e$\_$music$\_$remastering$\_$system}, and Jukebox \footnote{https://github.com/openai/jukebox}.
Feature vectors of each song were extracted with the dimension denoted in Table \ref{tab:data}.
Sixteen songs per batch were trained with CLEP, which has a network architecture of MLP with 4, 5, and 5 layers for CLMR, MEE, and Jukebox, respectively.
The preference prediction stage has a network architecture of 3-layer MLP.
Both CLEP and preference prediction stage were trained using Adam optimizer with learning rate scheduled so that it is reduced when validation loss is not decreasing until two epochs.
CLEP was trained for 20 epochs, and the learning rate was scheduled starting from 0.01.
The preference prediction stage was trained for 30 epochs with the learning rate starting from 0.001.

\begin{table*}[]
\caption{\textbf{Median values of accuracy, precision, recall, AUROC, and false positive rate (FPR) according to the musical feature extraction models and our models.
The reported $\chi^2$ values and their p-values are obtained with Friedman test (Statistical significance : *** $p < 0.001$, ** $p < 0.01$, * $p < 0.05$).}}
\begin{imageonly}
\resizebox{1.9\columnwidth}{!}{
\renewcommand{\arraystretch}{1.1}
\begin{tabular}{ccccccc}
\toprule
\textbf{Front-end Models} & \textbf{CLEP} & \textbf{Accuracy ($\uparrow$)} & \textbf{Precision ($\uparrow$)} & \textbf{Recall ($\uparrow$)} & \textbf{AUROC ($\uparrow$)} & \textbf{FPR ($\downarrow$)} \\ \hline
& CLEP-PN & 0.62 & 0.37 & 0.367 & 0.508 & 0.329 \\
& CLEP-P & 0.56  & 0.424 & \textbf{0.722} & \textbf{0.588} & 0.547 \\
CLMR & CLEP-N & \textbf{0.66} & \textbf{0.5} & 0.16 & 0.514 & \textbf{0.097} \\ \cdashline{2-7}[0.7pt/1.5pt]
& {$\chi^2$} (df=2) & \begin{tabular}[c]{@{}c@{}}9.621\\ (p=0.008**)\end{tabular}  & \begin{tabular}[c]{@{}c@{}}1.595\\ (p=0.451)\end{tabular} & \begin{tabular}[c]{@{}c@{}}25.613\\ (p=2.74e-06***)\end{tabular}  &  \begin{tabular}[c]{@{}c@{}}2.083\\ (p=0.353)\end{tabular} & \begin{tabular}[c]{@{}c@{}}26.547\\ (p=1.72e-06***)\end{tabular} \\ \hline
& CLEP-PN & 0.59 & 0.334 & 0.453 & 0.502 & 0.352 \\
& CLEP-P & 0.55 & 0.375 & \textbf{0.481} & 0.519 & 0.439 \\
MEE & CLEP-N & \textbf{0.61} & \textbf{0.404} & 0.367 & \textbf{0.538} & \textbf{0.286} \\ \cdashline{2-7}[0.7pt/1.5pt]
& {$\chi^2$} (df=2) & \begin{tabular}[c]{@{}c@{}}7.101\\ (p=0.029*)\end{tabular} & \begin{tabular}[c]{@{}c@{}}1.916\\ (p=0.384)\end{tabular} & \begin{tabular}[c]{@{}c@{}}15.475\\ (p=0.0004***)\end{tabular} & \begin{tabular}[c]{@{}c@{}}1\\ (p=0.607)\end{tabular} &  \begin{tabular}[c]{@{}c@{}}20.609\\ (p=3.35e-05***)\end{tabular}     \\ \hline
& CLEP-PN & 0.59 & 0.421 & 0.547 & 0.5 & 0.423 \\
& CLEP-P & 0.64  & 0.457 & \textbf{0.747} & \textbf{0.653} & 0.431 \\
Jukebox & CLEP-N & \textbf{0.7} & \textbf{0.519} & 0.338 & 0.555 & \textbf{0.15} \\ \cdashline{2-7}[0.7pt/1.5pt]
& {$\chi^2$} (df=2)     & \begin{tabular}[c]{@{}c@{}}11.5\\ (p=0.003**)\end{tabular}  & \begin{tabular}[c]{@{}c@{}}18.583\\ (p=9.22e-05***)\end{tabular} & \begin{tabular}[c]{@{}c@{}}16.28\\ (p=0.0003***)\end{tabular} & \begin{tabular}[c]{@{}c@{}}21.894\\ (p=1.76e-05***)\end{tabular} & \begin{tabular}[c]{@{}c@{}}25.872\\ (p=2.41e-06***)\end{tabular} \\ \bottomrule
\end{tabular}
}
\end{imageonly}
\label{tab:result}
\end{table*}

\subsection{User Preference Dataset}

We conducted a web-based survey asking participants about their music preferences to train and evaluate our models.
It is difficult to define music preferences elaborately, but as many online music recommendation services do, user feedback can be elicited to assume their preferences ~\cite{jawaheer2010comparison}.
The survey asked for the likes and dislikes of certain songs, and the collected data were used to represent each participant's music preference.

Twenty-four volunteers with no hearing problems were recruited from online student communities.
They were all Koreans, and their ages ranged from 24 to 37, with an average of 27.
After briefly introducing the survey process, we obtained consent for their participation.
They were asked to listen to 200 music clips and answer whether they liked or disliked each song.
Since users' familiarity with songs does affect their preference ~\cite{peretz1998exposure}, 40 songs were randomly selected from different genres to reduce genre bias and effects on the popularity of the songs.
They consisted of the five most popular genres nowadays - rock, EDM, hip-hop, pop, and R\&B.
We used the 'Get Recommendations' function provided in Spotify API \footnote{https://developer.spotify.com/documentation/web-api/}, which can return a list of tracks when given a particular genre.
Music excerpts of 10 seconds were randomly selected from each track and given in random order to each participant.
By referring to ~\cite{peretz1998exposure}, which studied music preference and recognition, it was considered that 10 seconds were enough for the participants to identify the melodies and decide their preferences on each song.
The music clips were given stereo-channeled with a sampling rate of 44.1kHz in the survey but were manipulated in the feature extraction stage to fit each feature extraction model.

Through the survey, we obtained each participant's preferences for 200 songs.
Each participant had a different ratio of their liked and disliked songs - some had much more liked songs while some had much more disliked songs.
The average ratio of the number of liked songs to the number of disliked songs was 0.96:1 on average, saying the preferences of the entire participants were not biased.
Within the 200 individual data, we divided them into a training set and a test set at a ratio of 3:1.
We then trained the models with the training set and assessed their performances on the test set.

\begin{table*}[]
\caption{\textbf{P-values of Wilcoxon signed-rank test as a post-hoc analysis of the Friedman test above (Statistical significance : *** $p < 0.001$, ** $p < 0.01$, * $p < 0.05$).
Significant order relations between the models are noted on the right side.}}
\resizebox{1.6\columnwidth}{!}{
\renewcommand{\arraystretch}{1.2}
\begin{tabular}{cccccc}
\toprule
& & \textbf{CLEP-PN vs P} & \textbf{CLEP-P vs N} & \textbf{CLEP-N vs PN} & \textbf{Results} \\ \hline
& CLMR & 0.065  & 0.002** & 0.028* & CLEP-N > PN, P \\
Accuracy ($\uparrow$) & MEE & 0.648 & 0.004** & 0.016* & CLEP-N > PN, P  \\
& Jukebox & 0.016* & 0.038* & 0.008** & CLEP-N > P > PN \\ \hline
& CLMR & 0.031* & 9.6e-05*** & 0.009** & CLEP-P > PN > N \\
Recall ($\uparrow$) & MEE  & 0.298 & 0.0003*** & 0.029*& CLEP-PN, P > N\\
& Jukebox & 0.026* & 3.6e-05*** & 0.042* & CLEP-P > PN > N \\ \hline
& CLMR & 0.066 & 7.6e-05*** & 0.001** & CLEP-N < PN, P \\
FPR ($\downarrow$) & MEE  & 0.173 & 0.0003*** & 0.03* & CLEP-N < PN, P \\
& Jukebox & 0.82 & 1.9e-05*** & 0.001** & CLEP-N < PN, P \\ \bottomrule
\end{tabular}
}
\label{tab:posthoc}
\end{table*}

\begin{figure*}
    \caption{\textbf{Example of t-SNE visualization of embedding spaces trained with data obtained from a single participant, with MEE as musical feature extractor. Red points represent the songs with positive preference, and blue points represent the songs with negative preference.}}
    \includegraphics[width=1.8\columnwidth]{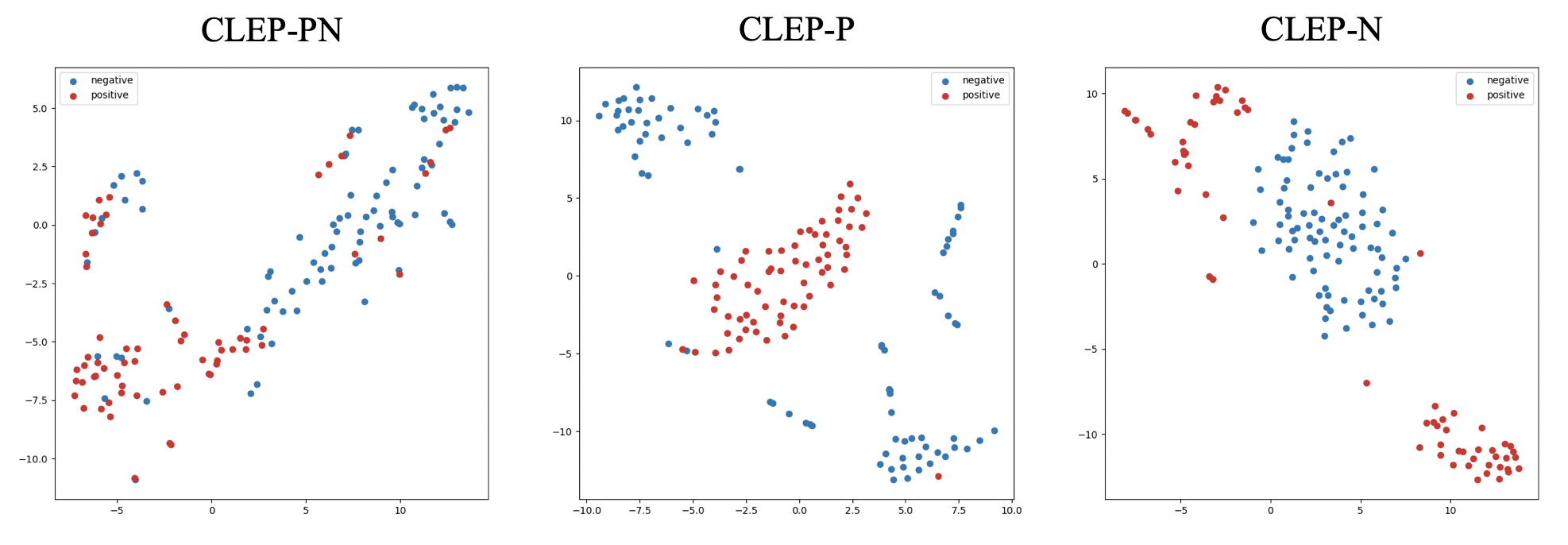}
    \label{fig:embed}
    \vspace{10pt}
\end{figure*}

\subsection{Evaluation}

To compare the performance of each model, we used the following five metrics - accuracy, precision, recall, area under the receiver operating characteristic curve (AUROC), and false positive rate.
Accuracy, precision, recall, and AUROC are the standard metrics for evaluating recommendation systems.
In addition, we also measured the false positive rate, which is a ratio of misclassified negatives to the total negatives.
The recommendation system field has plentiful evaluation methodologies ~\cite{bellogin2011precision, herlocker2004evaluating, shani2011evaluating}, but most of them focus on true positives as the evaluation objective.
However, ~\cite{mena2020agreement} points out that false positives are a clear concern in music recommendations.
From the user experience perspective, users are not aware of not being recommended a song they like.
Instead, it is more disappointing to be recommended a song they dislike.
Since false positives negatively affect user experience compared to false negatives, measuring false positive metrics will help analyze the practical utility of a music recommendation system.
From this point of view, it is crucial to look into precision and false positive rate, which are the metrics relevant to false positives.
Precision in recommendation refers to the ratio of liked songs over recommended ones.
Meanwhile, the false positive rate is the ratio of recommended songs over the songs that which user truly dislikes.
High precision and low false positive rate imply that the recommendation system is worthwhile in terms of user experience.

The experimental results were analyzed through the Friedman test, and the overall test results are illustrated in Table \ref{tab:result}.
The following $\chi^2$ and p-value in the table demonstrate the statistical significance that the results differ by model.
The results of different musical feature extractions are also displayed in the table, showing a consistent tendency to some degree regardless of feature extractors.
Accuracy, recall, and false positive rate showed statistically significant differences in all three cases, while precision and AUROC showed differences only in models using Jukebox for its feature extractor.
In order to verify specified relationships between the models, Wilcoxon signed-rank tests were performed as a post-hoc analysis for accuracy, recall, and false positive rate.
Table \ref{tab:posthoc} shows multiple testing results between the models.

\section{Results and Discussion}

We have trained our three models - CLEP-PN, CLEP-P, and CLEP-N - to embed the contents of songs exploiting preferences and predict the preference of unknown songs. 
In the training phase, each data was embedded depending on its feature and preference.
We observed that the songs were embedded as expected when visualized in two dimensions using t-SNE as is seen in Figure \ref{fig:embed} for instance.
Songs with positive and negative preferences were clustered each in the embedding space of CLEP-PN.
Furthermore, songs with positive preferences were clustered while songs with negative preferences were spread out in the embedding space of CLEP-P, and vice versa in the case of CLEP-N.

As we showed, there were statistically significant differences between the models in terms of accuracy, recall, and false positive rate.
First, CLEP-N showed the highest accuracy among the three models.
Although the statistical significance for the difference between CLEP-PN and CLEP-P was slightly different depending on the musical feature extractors, the accuracy of CLEP-N consistently exceeded the accuracy of the other two models.
In the case of precision, models which used Jukebox as its musical feature extractor only showed a significant difference ($\chi^2(2)=18.583, p<0.001$), presenting the highest value in CLEP-N.
The precision of models with other musical feature extractors showed a lack of significance, but the median values were consistently the highest in CLEP-N.

Meanwhile, the results showed that CLEP-N performed the lowest recall.
In the case of false positive rate, CLEP-N outperformed the other two regardless of musical feature extractors, showing the lowest value.
A recommendation system's low false positive rate implies that the model barely recommends music the user dislikes.
The results of CLEP-N showing high false positive rate and low recall indicate that it is better at predicting songs the user dislikes than predicting songs the user likes.
A simple approach can allow a rough guess of thinking that CLEP-N is too pessimistic, predicting that the user dislikes every song.
However, considering that the survey data was balanced in terms of preferences and CLEP-N showed the highest accuracy, it is convincing enough to claim its strength.
The false positive rate and recall both have actual preferences as the denominator, but the false positive rate focuses on the negatives while recall focuses on the positives.
False positive rate is an anti-metric for recall, which is a metric aware of the irrelevant items returned by the recommendation systems.
Considering that anti-metrics are more valuable than classical metrics when distinguishing recommendation systems with similar relevance ~\cite{sanchez2018measuring}, the fact that CLEP-N is showing a high false positive rate is strong evidence of its potential to be utilized in recommendation systems.

All three models showed no particular tendency in terms of AUROC, and the values were insufficient to state the stable performance of each model.
As seen from the low AUROC, our models, including CLEP-N, have limitations in their immediate application as a recommendation system.
It is due to the shortage of data in quantity and the simple implementation aimed at identifying the differences, and adjusting CLEP-N for real application will be left as our future work.

Based on the results, the research questions of our work as mentioned above can be discussed like the following:

\begin{itemize}
    \item \textbf{RQ 1. What characteristics do negative preferences have in terms of explaining music taste?}

If we think of a user's music taste as a complex distribution of songs the user likes and dislikes, we were interested in which of these three models most similarly simulates the distribution.
If the contents of songs that the user feels positive or negative have a certain tendency, the features of the songs with the same preference will be embedded close to each other.
Thus we can regard the embedding spaces of our models as the distribution of users' music tastes according to their positive and negative preferences.
Based on the result that CLEP-N showed the highest accuracy, we provide evidence that songs with negative preference have more distinct characteristics than songs with positive preference.
It is also supported by the concept of serendipity ~\cite{kotkov2016survey}, which is a measure indicating the unexpectedness of a recommendation.
The fact that users react to unexpectedly good things points out that there is a chance of finding songs the user may like in an unpredictable area of the user's music taste, and the findings of our work explain it.\\

    \item \textbf{RQ 2. How does applying negative preference help improve music recommendations?}

Although our experimental settings had a gap from the real-world situation, we verified the model's potential to exploit negative preference in content-based music recommendations by conditioning the preferences in the models. 
From the perspective of user experience, it is shown that the model with a low false positive rate and high precision can lead the users to a pleasant experience of consuming music. 
Through our work, we verified that CLEP-N showed a distinctly low false positive rate and, in some cases, high precision.
Therefore, we can conclude that exploiting negative preference contributes to improvement in false positive metrics, and this consideration in music recommendations will be expected to make significant progress.\\
\end{itemize}

\section{Conclusion} 

In this work, we analyzed the role of negative preferences in users' music tastes by comparing three models with differently conditioned contrastive learning exploiting preference (CLEP) - models exploiting both positive and negative preferences (CLEP-PN), positive preference only (CLEP-P), and negative preference only (CLEP-N).
We found that CLEP-N, which assumes that negative preference is more characterized, showed the highest accuracy among the three proposed models.
It leads to a conclusion that negative preference has the potential to have more explainable characteristics in users' music taste compared to positive preference.
Furthermore, CLEP-N outperformed the other two models in terms of false positive metrics.
As false positive metrics are told as highly relevant in recommendation literature, CLEP-N also illuminates the capacity of improving music recommendations by utilizing negative preferences.

We have intensified our work to enlighten the effects of negative preference through comparative analysis.
In other words, our work is focused on synthesizing our novel findings for negative preferences but not on directly applicable model proposals.
In future work, we plan to investigate practical ways to exploit our findings in real-world applications.
Moreover, although the experiment was designed in a binary manner to compare the characteristics of the two extremes of preferences, we also intend to develop this study in-depth by redesigning the experiment to inquire about music preferences quantitatively.

\begin{acks}
We would like to thank Junghyun Koo for his assistance and insightful comments.
This work was supported partly by Institute of Information \& communications Technology Planning \& Evaluation (IITP) grant funded by the Korea government (MSIT) (No. 2022-0-00320) and partly by Culture, Sports and Tourism R\&D Program through the Korea Creative Content Agency grant funded by the Ministry of Culture, Sports and Tourism in 2022 (No. R2022020066).
\end{acks}

\balance
\bibliographystyle{ACM-Reference-Format}
\bibliography{main}

\end{document}